\documentclass[10pt,twocolumn,letterpaper]{article}

\usepackage{cvpr}
\usepackage{times}
\usepackage{epsfig}
\usepackage{graphicx}
\usepackage{amsmath}
\usepackage{amssymb}
\usepackage{subcaption, xcolor}

\usepackage[pagebackref=true,breaklinks=true,letterpaper=true,colorlinks,bookmarks=false]{hyperref}

\cvprfinalcopy 

\def\cvprPaperID{3768} 

\ifcvprfinal\fi
\begin{document}

\title{Natural and Realistic Single Image Super-Resolution with Explicit Natural Manifold Discrimination}

\author{Jae Woong Soh \qquad Gu Yong Park \qquad Junho Jo \qquad Nam Ik Cho\\
Department of ECE, INMC, Seoul National University, Seoul, Korea\\
{\tt\small \{soh90815, benkay, jottue\}@ispl.snu.ac.kr, nicho@snu.ac.kr}
}
\maketitle
\thispagestyle{empty}

\begin{abstract}
Recently, many convolutional neural networks for single image super-resolution (SISR) have been proposed, which focus on reconstructing the high-resolution images in terms of objective distortion measures. However, the networks trained with objective loss functions generally fail to reconstruct the realistic fine textures and details that are essential for better perceptual quality. Recovering the realistic details remains a challenging problem, and only a few works have been proposed which aim at increasing the perceptual quality by generating enhanced textures. However, the generated fake details often make undesirable artifacts and the overall image looks somewhat unnatural.
Therefore, in this paper, we present a new approach to reconstructing realistic super-resolved images with high perceptual quality, while maintaining the naturalness of the result. In particular, we focus on the domain prior properties of SISR problem. Specifically, we define the naturalness prior in the low-level domain and constrain the output image in the natural manifold, which eventually generates more natural and realistic images. Our results show better naturalness compared to the recent super-resolution algorithms including perception-oriented ones.
\end{abstract}

\begin{figure}[t]
	\begin{center}
	
		\begin{subfigure}[t]{0.48\linewidth}
		\centering
		\includegraphics[width=1\columnwidth]{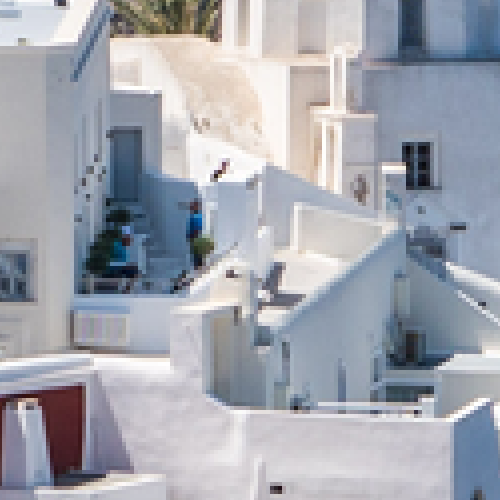}
		\caption{HR}
	\end{subfigure}
	\begin{subfigure}[t]{0.48\linewidth}
		\centering
		\includegraphics[width=1\columnwidth]{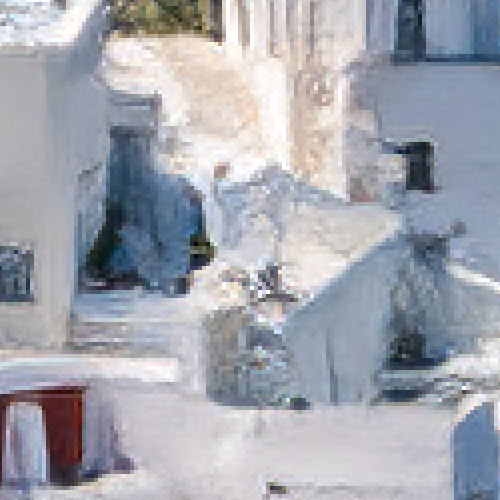}
		\caption{EnhanceNet \cite{EnhanceNet}}
	\end{subfigure}
	\begin{subfigure}[t]{0.48\linewidth}
	\centering
	\includegraphics[width=1\columnwidth]{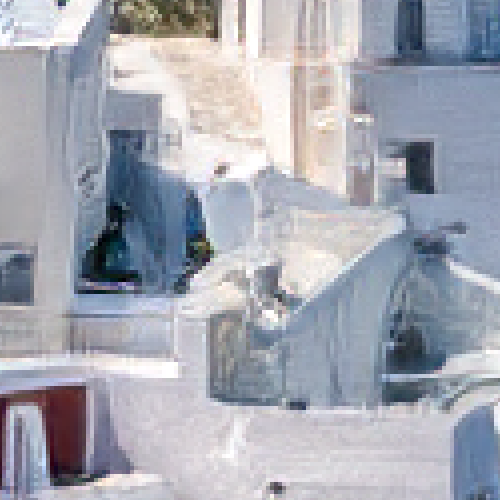}
	\caption{SFT-GAN \cite{SFT-GAN}}
\end{subfigure}
\begin{subfigure}[t]{0.48\linewidth}
	\centering
	\includegraphics[width=1\columnwidth]{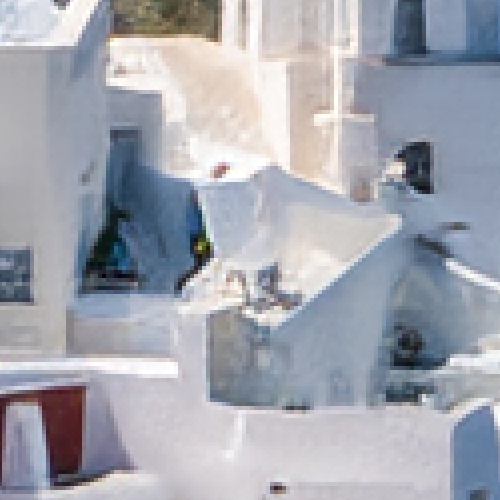}
	\caption{Our NatSR}
\end{subfigure}

	\end{center}
	\caption{Super-resolved results ($\times 4$) of ``0823'' in DIV2K validation set \cite{NTIRE}. A part of the image is cropped and zoomed for visualization. Our NatSR result is more natural with less artifacts which is perceptually plausible than other algorithms' results.}
	\label{fig:001}
\end{figure}

\section{Introduction}

Single image super-resolution (SISR) is a classical image restoration problem which aims to recover a high-resolution (HR) image from the corresponding low-resolution (LR) image. In SISR problems, the given image is usually assumed to be a low-pass filtered and downsampled version of an HR image. Hence, recovering the HR is an ill-posed problem since multiple HR images can correspond to one LR image. That is, the SISR is a challenging one-to-many problem which attracted researchers to find many interesting solutions and applications, and thus numerous algorithms have been proposed so far.

Recently, convolutional neural networks (CNNs) have shown great success in most computer vision areas including the SISR. In typical CNN-based SISR methods, the distortion-oriented loss functions are considered. Specifically, the CNNs attempt to achieve higher peak-signal-to-noise ratio (PSNR), \textit{i.e.}, low distortion in terms of mean squared error (MSE). There have been lots of distortion-oriented CNNs for SISR~\cite{SRCNN, VDSR, ESPCN, DRCN, LapSRN, DRRN, SRDenseNet, EDSR, RDN, IDN, DSRN}, and the performance of SISR is ever increasing as many researchers are still creating innovative architectures and also as the possible depth and connections of the networks are growing.
However, they yield somewhat blurry results and do not recover the fine details even with very deep and complex networks. It is because the distortion-oriented models' results are the average of possible HR images.

To resolve the above-stated issues, perception-oriented models have also been proposed for obtaining better perceptual quality HR images. For some examples, the perceptual loss was introduced in \cite{Perceptual}, which is defined as the distance in the feature domain. More recently, SRGAN~\cite{SRGAN} and EnhanceNet~\cite{EnhanceNet} have been proposed for producing better perceptual quality. The SRGAN employed generative models, particularly the generative adversarial nets (GAN)~\cite{GAN}, and adopted the perceptual loss. The EnhanceNet added an additional texture loss~\cite{Texture} for better texture reconstruction. However, they sometimes generate unpleasant and unnatural artifacts along with the reconstructed details.

There have also been some methods that consider the naturalness of super-resolved images.
One of these approaches is to implicitly supervise the naturalness through the refined dataset. Specifically, as the CNN is very sensitive to the training dataset, several methods \cite{EDSR, RDN} considered using the refined dataset. For example, patches with low gradient magnitudes are discarded from the training dataset, which provides better naturalness implicitly. This approach might increase the PSNR performance by constraining the possible HR space to the rich-textured one.
Another approach is to provide explicit supervision by conditioning the feature spaces. For example, the recently developed SFT-GAN \cite{SFT-GAN} has shown great perceptual quality by constraining the features with its high-level semantics while adopting the adversarial loss. However, its practical usage is limited because it requires the categorical prior, and also it is limited to the categories which are included in the training process. For the out-of-category inputs, this framework is the same as SRGAN \cite{SRGAN}. Moreover, SFT-GAN strongly relies on the ability of the adopted semantic segmentation method because the wrong designation of semantics might cause worse perceptual quality.

For obtaining realistic and natural perceptual quality HR images, we propose a new SISR approach which constrains the low-level domain prior instead of high-level semantics. For this, we first investigate the process and the domain knowledge of SISR. By exploiting the domain knowledge, we explicitly model the HR space of corresponding LR image, and build a discriminator which determines the decision boundary between the natural manifold and unnatural manifold. By constraining the output image into the natural manifold, our generative model can target only one of the multi-modal outputs in the desired target space. As a results, our method shows less artifacts than other perception-oriented methods as shown in Figure~\ref{fig:001}.

In summary, the main contributions of this paper are as follows.
\begin{itemize}
    \item We model the SISR problem explicitly and investigate the desirable HR space.
    \item We design a CNN-based natural manifold discriminator and show our model is reasonable.
    \item We adopt a CNN structure with fractal residual learning (FRL) and demonstrate a distortion-oriented model named fractal residual super-resolution (FRSR), which achieves comparable results to recent CNNs.
    \item We propose a perception-oriented SISR method named as natural and realistic super-resolution (NatSR), which generates realistic textures and natural details effectively while achieving high perceptual quality.
\end{itemize}

The rest of this paper is organized as follows. In Sec. \ref{sec:SISR}, we explicitly model the LR-HR space and the SISR problem, and investigate its inherent properties. Then in Sec. \ref{sec:DIS}, we divide the target HR space into three disjoint sets where two sets are in the unnatural manifold and the one is in the natural manifold. In Sec. \ref{sec:MODEL}, we demonstrate our main method and the NatSR, and in Sec. \ref{fig:analysis} we discuss and analyze the feasibility in several ways. The experimental results are shown in Sec. \ref{sec:Results}.


\section{Related Work}

\subsection{Single Image Super-Resolution}
The conventional non-CNN methods mainly focused on the domain and feature priors. Early methods explored the domain priors to predict missing pixels. For example, interpolation methods such as bicubic and Lanczos generate the HR pixels by the weighted average of neighboring LR pixels. Later, the priors such as edge feature, gradient feature~\cite{Prior1, Prior2} and internal non-local similarity \cite{Self-Exemplar} were investigated. Also, dictionary learning sparse coding methods were exploited for the SISR~\cite{SparseSR, EmbedSR, CoupleSR, A+}.
Recently, it has been shown that CNN-based methods outperform the earlier non-CNN algorithms, showing great breakthrough in accuracy. These CNN-based methods implicitly adopt image and domain priors which are inscribed in training datasets. The SRCNN~\cite{SRCNN} was the first CNN-based method which uses three convolution layers, and many other works with deeper and heavier structure have been proposed afterward~\cite{VDSR, ESPCN, DRCN, DRRN, LapSRN, SRDenseNet, EDSR, RDN, IDN, DSRN}. All these methods are discriminative and distortion-oriented approaches, which aim to achieve higher PSNR.

\subsection{Perception Oriented Super-Resolution}
The problem of distortion-oriented models recently drew the attention of researchers that the super-resolved results often lack the high-frequency details and are not perceptually satisfying. Also, Blau \etal~\cite{perception} showed that there is a trade-off between the perceptual quality and distortion, and some perception-oriented models have been proposed accordingly. For example, Johnson \etal \cite{Perceptual} have shown that the loss in the pixel domain is not optimal for the perceptual quality, and instead, the loss in the feature space might be closer to the human perception model. Then, Ledig \etal \cite{SRGAN} introduced the SRGAN which adopted the generative model with GAN~\cite{GAN} and employed the perceptual loss as in~\cite{Perceptual}. Hence, unlike the distortion-oriented methods that produce the average of possible HR images, the SRGAN generates one of the candidates in the multi-modal target HR space. EnhanceNet \cite{EnhanceNet} goes one step further by exploiting the texture loss~\cite{Texture} for better producing image details. However, due to the inherent property of one-to-many inverse problem, it is required to consider the semantics for the generated pixels. In this respect, SFT-GAN~\cite{SFT-GAN} restricts the feature space by conditioning the semantic categories of target pixels.

\section{Modeling the SISR}
\label{sec:SISR}

\begin{figure}[t]
	\begin{center}
		\begin{subfigure}[b]{0.45\linewidth}
			\centering
			\includegraphics[width=1\columnwidth]{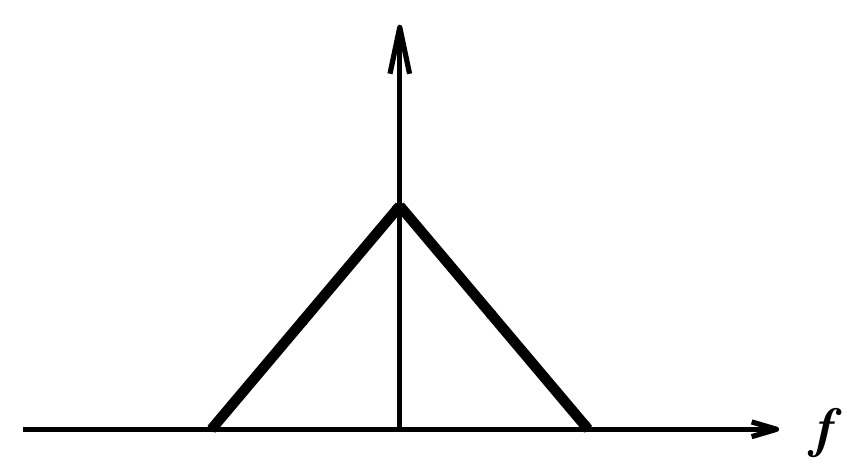}
			\caption{HR}
				\label{fig:HR}
		\end{subfigure}
		\begin{subfigure}[b]{0.45\linewidth}
		\centering
		\includegraphics[width=1\columnwidth]{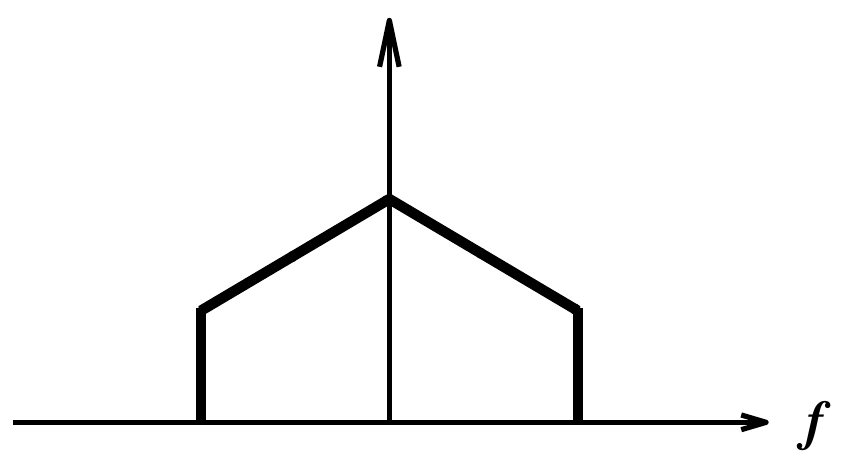}
		\caption{LR}
			\label{fig:LR}
	\end{subfigure}
		\begin{subfigure}[b]{0.45\linewidth}	
		\centering
		\includegraphics[width=1\columnwidth]{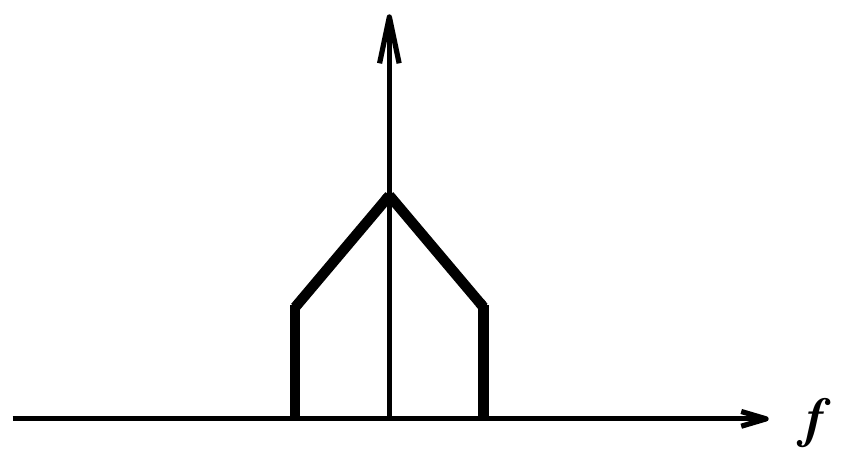}
		\caption{Blurry HR}
			\label{fig:BLURRY}
	\end{subfigure}
		\begin{subfigure}[b]{0.45\linewidth}
			\centering
		\includegraphics[width=1\columnwidth]{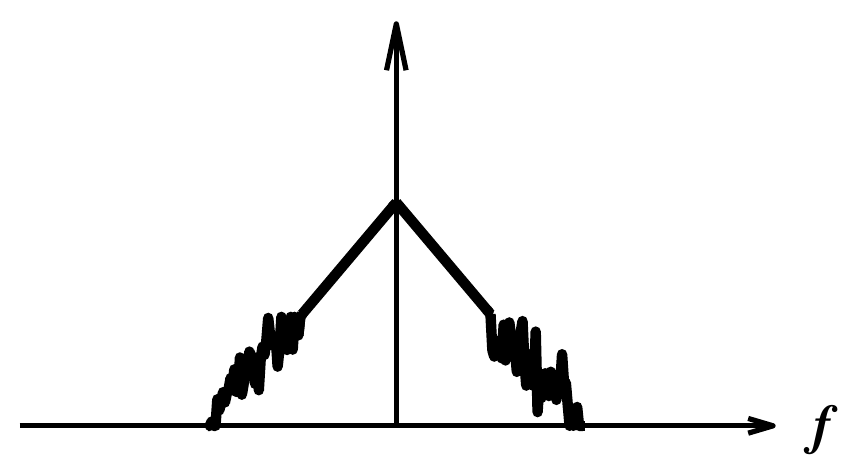}
		\caption{Noisy HR}
			\label{fig:NOISY}
	\end{subfigure}

	\end{center}
	\caption{A simple explanation of LR-HR relationship and SISR in the frequency domain.}
	\label{fig:FREQ}
\end{figure}

In this section, we explicitly define and model the LR-HR space and the SISR problem. First of all, let us define the LR image $I_{LR}$ as the low-pass filtered and downsampled HR image $I_{HR}$. Formally, the LR-HR relation is described as 
\begin{equation}
I_{LR}=h(I_{HR})^\downarrow,
\label{HRtoLR}
\end{equation}
where $h(\cdot)$ denotes a low-pass filter and $\downarrow$ denotes downsampling. 
\figurename~{\ref{fig:HR}} and \figurename{~\ref{fig:LR} show a simple explanation of HR and LR correspondence in the frequency domain where we assume that the spatial domain is infinite. Both \figurename{~\ref{fig:BLURRY}} and \figurename{~\ref{fig:NOISY}} are possible HRs for the corresponding LR in \figurename~{\ref{fig:LR}}, and moreover, there can be infinite number of possible HRs that have the same low frequency components but different high-frequency parts (denoted noisy in \figurename{~\ref{fig:NOISY}}). 
As the SISR is to find an HR for the given LR, it is usually modeled as finding the conditional likelihood $p(I_{HR}|I_{LR})$. Due to its one-to-many property, it is better to model it as a generative model rather than a discriminative one.

\section{Natural Manifold Discrimination}
\label{sec:DIS}
\subsection{Designing Natural Manifold}

\begin{figure}[t]
	\begin{center}
		\includegraphics[width=0.8\linewidth]{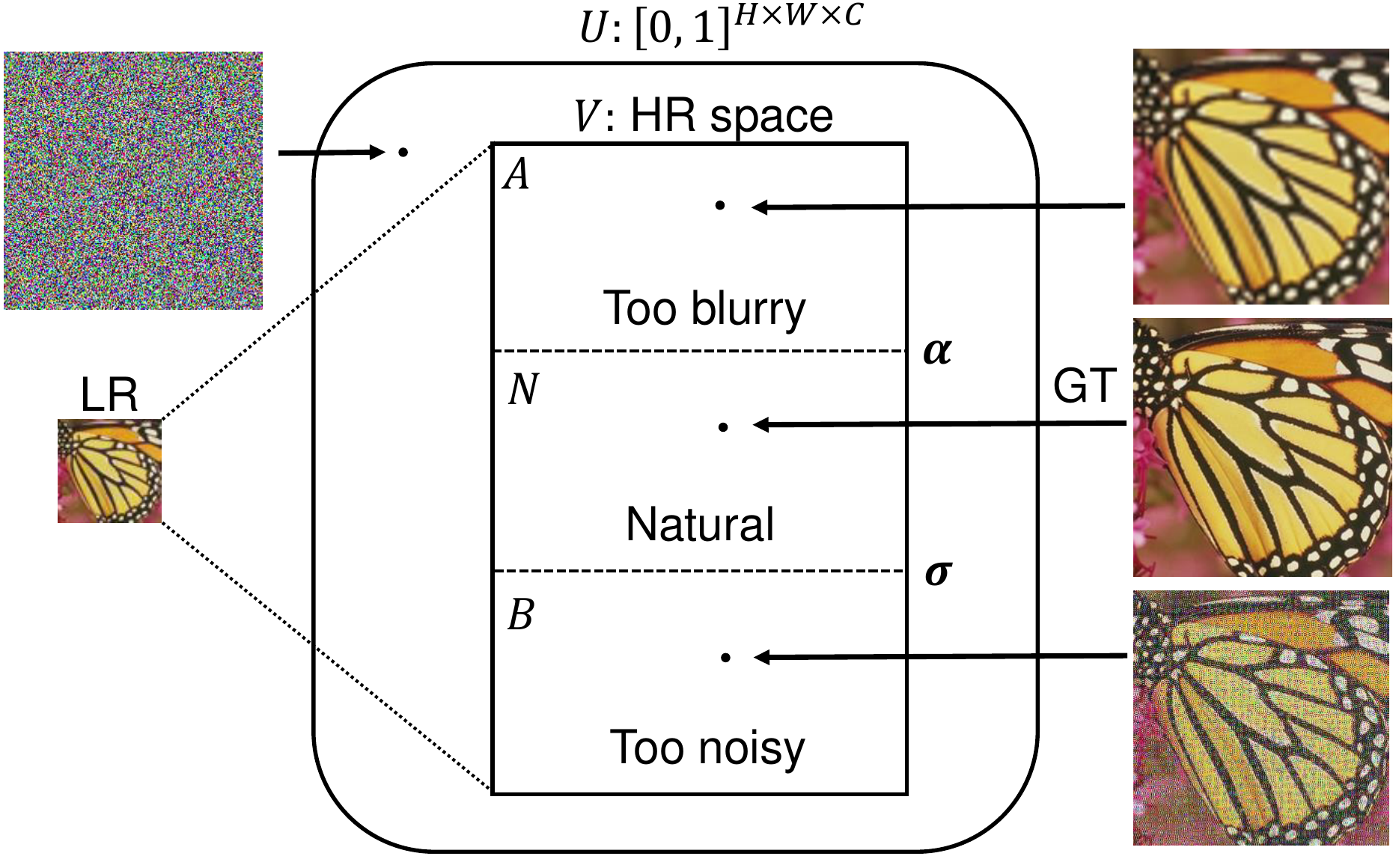}
	\end{center}
	\caption{Our proposed LR-HR model of the natural manifold and its discrimination for SISR. $U$ is the image space, $V$ is the possible HR space, and $A, B,$ and $N$ are three disjoint sets of $V$. $\alpha$ and $\sigma$ control the boundary between the manifolds.}
	\label{fig:manifold}
\end{figure}

We now go into the real situation to find the natural manifold. \figurename{~\ref{fig:manifold}} shows our LR-HR image space modeling, where $U: [0,1]^{H\times W\times C}$ is the overall image set with height $H$, width $W$, and channel $C$ with the normalized pixel value. For a certain $I_{LR}$, $V$ is the space whose elements all results into the same $I_{LR}$ by the low-pass filtering and downsampling. Conversely, an LR image is mapped to an element in $V$ by any SR method.
We may also interpret the early CNNs with our LR-HR model. For the distortion-oriented models, the output is the average of the elements in the HR space, \textit{i.e.}, $\sum w_i I_{HR_i}$ where $I_{HR_i} \in V$, for some $i$ and weights $w_i$, and thus the result is blurry. To alleviate this problem, some methods \cite{EDSR, RDN} refined the training set. Specifically, they discarded the training patches with low gradient magnitudes, which gives implicit constraints on the candidate $I_{HR_i}$'s to keep the resulting outputs away from the blurry images.

To model the natural manifold, we divide $V$ into three disjoint sets as illustrated in \figurename{~\ref{fig:manifold}}. The first one is the blurry set $A$, the elements of which are modeled as the convex combination of interpolated LR and the original HR. Specifically, the set $A$ is defined as
\begin{equation}
A=\{I_{A}|I_{A}=(1-\alpha) h(I_{LR}^\uparrow) + \alpha I_{HR}\},
\label{LRtoHR}
\end{equation}
where $h(\cdot)$ is the same low-pass filter as in eq.~(\ref{HRtoLR}), and $\uparrow$ denotes upsampling with zero insertion between original values. Hence, $h(I_{LR}^\uparrow)$ corresponds to \figurename{~\ref{fig:BLURRY}} which also means the interpolation of $I_{LR}$ to the size of $I_{HR}$. Also, the $\alpha \in [0, 1]$ is a hyper-parameter which decides the decision boundary between the set $A$ and $N$, {\em i.e.}, between the \figurename{~\ref{fig:BLURRY}} and \figurename{~\ref{fig:HR}}.
We can easily show that the $I_{A}$ defined above is also an element of $V$, {\em i.e.}, $A \subset V$. To be specific, if we apply low-pass filtering and downsampling to the $I_{A}$, it becomes an LR as follows: 
\begin{align}
&h(I_{A})^\downarrow\\
=&h((1-\alpha)h(I_{LR}^\uparrow)+\alpha I_{HR})^\downarrow\\
=&h((1-\alpha)h(I_{LR}^\uparrow))^\downarrow + h(\alpha I_{HR})^\downarrow\\
=&(1-\alpha) h(I_{LR}^\uparrow)^\downarrow + \alpha h(I_{HR})^\downarrow\\
=&(1-\alpha) I_{LR} + \alpha I_{LR}\\
=&I_{LR}.
\end{align}
Hence, from eq.(\ref{HRtoLR}), it is shown that $I_{A} \in V$. In other words, the weighted sum of \figurename{~\ref{fig:BLURRY}} and \figurename{~\ref{fig:HR}} is of course in the $V$.

The second set to consider is the noisy set $B$, which contains the images like \figurename{~\ref{fig:NOISY}}. 
Specifically, we can model the set as:
\begin{equation}
B=\{I_{B}|I_{B}=I_{HR}+n\}
\label{noisy_image}
\end{equation}
where $n$ is the noise in the high-frequency, with standard deviation $\sigma$.
We can also see that $B \subset V$, because
\begin{align}
&h(I_{B})^\downarrow\\
=&h(I_{HR}+n)^\downarrow\\
=&h(I_{HR})^\downarrow+h(n)^\downarrow\\
=&h(I_{HR})^\downarrow\\
=&I_{LR}.
\end{align}
Also, $I_{B}$ can be interpreted as the convex combination of $I_{HR}$ and $I_{HR}+n_0$ (weighted sum of
\figurename{~\ref{fig:HR}} and \figurename{~\ref{fig:NOISY}}), because
\begin{align}
&(1-\beta)I_{HR} + \beta (I_{HR}+n_0)\\
=&I_{HR}-\beta I_{HR} + \beta I_{HR} + \beta n_0\\
=&I_{HR} + \beta n_0.
\end{align}
where $n = \beta n_0$.

\begin{figure}[t]
	\begin{center}
		\begin{subfigure}[b]{0.45\linewidth}
			\centering
			\includegraphics[width=1\columnwidth]{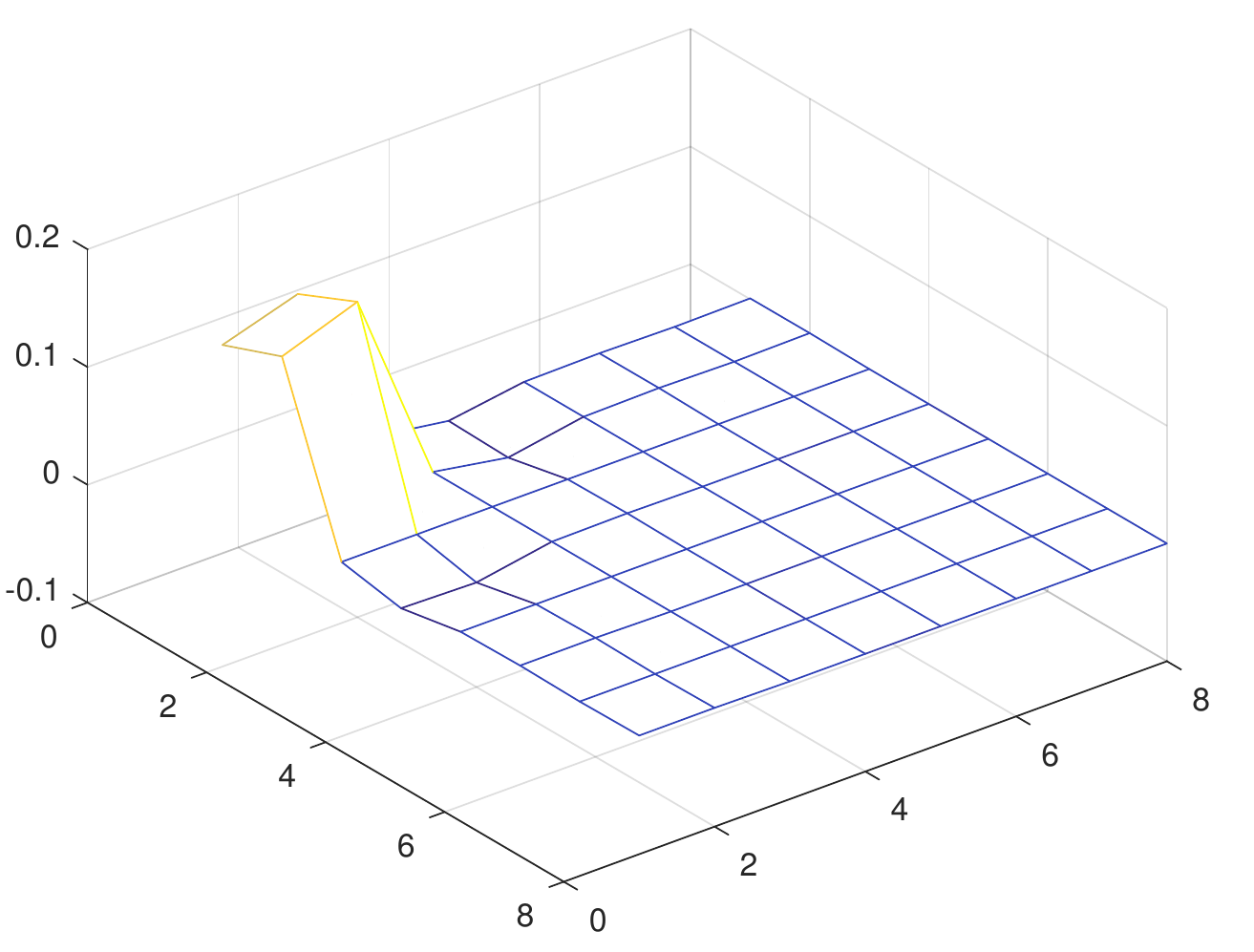}
			\caption{$8 \times 8$ DCT.}
			\label{fig:DCT8}
		\end{subfigure}
		\begin{subfigure}[b]{0.45\linewidth}
			\centering
			\includegraphics[width=1\columnwidth]{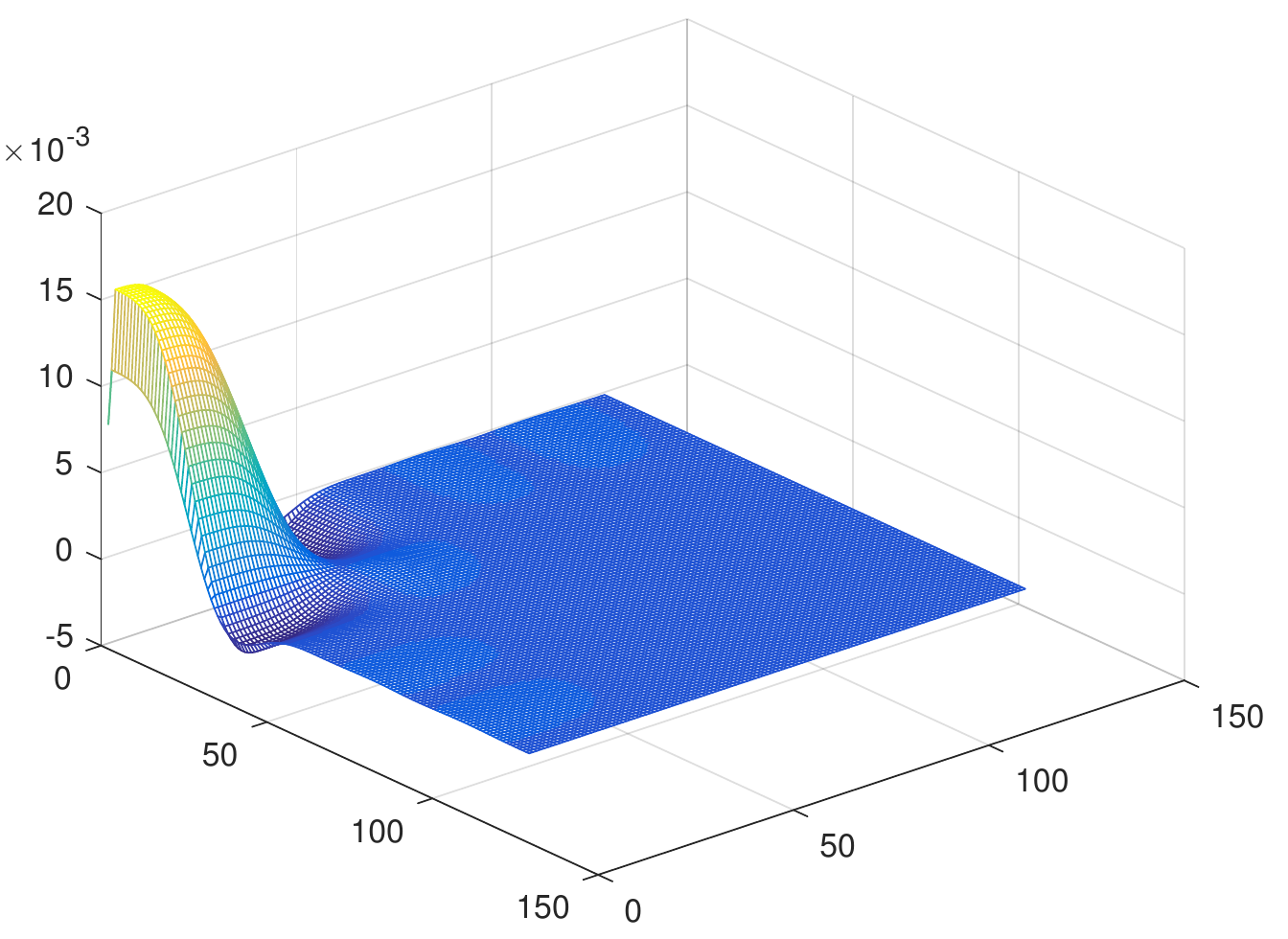}
			\caption{$128 \times 128$ DCT.}
			\label{fig:DCT128}
		\end{subfigure}

	\end{center}
	\caption{DCT coefficients of bicubic up/downsampling kernels for the scaling factor of $\times4$.}
	\label{fig:kernel}
\end{figure}
The blurry $I_A$ and noisy $I_B$ are used for training our natural manifold discriminator that will be explained in the next subsection.
In practice, we perform the noise injection in the frequency domain using 2D-discrete cosine transform (DCT). We set the low-pass filter for up/downsampling in eq.(\ref{HRtoLR}) and eq.(\ref{LRtoHR}) as the bicubic filter, and its DCT is shown in \figurename{~\ref{fig:kernel}}. To generate a wide range of noisy images, we inject the noise into the last column and row. In the experiments, we use the $8 \times 8$ 2D-DCT for brevity.

\subsection{Natural Manifold Discriminator}
To narrow the target space to the natural manifold, we design a discriminator that differentiates the natural image (the elements that belong to $N$ as in \figurename{~\ref{fig:manifold}}) from the blurry/noisy ones ($A$ or $B$). For this, we design a CNN-based classifier that discriminates $N$ (natural manifold) and $A \cup B$ (unnatural manifold), which will be called natural manifold discriminator (NMD). The training is performed with the sigmoid binary cross entropy loss function defined as
\begin{equation}
-\mathbb{E}_{x \in A\cup B}[\log (1-D_{NM}(x))] - \mathbb{E}_{x \in N}[\log (D_{NM}(x))],
\label{eq:nmd}
\end{equation}
where $D_{NM}(\cdot)$ denotes the output sigmoid value of NMD. For the expectation, we use the empirical mean of the training dataset. The network architecture of our NMD is shown in \figurename{~\ref{fig:nmd}}, which is a simple VGG-style CNN. Fully-connected layers for the last stage is not used in our case. Instead, one convolution layer and global average pooling are used.
\begin{figure}[t]
	\begin{center}
		\includegraphics[width=0.99\linewidth]{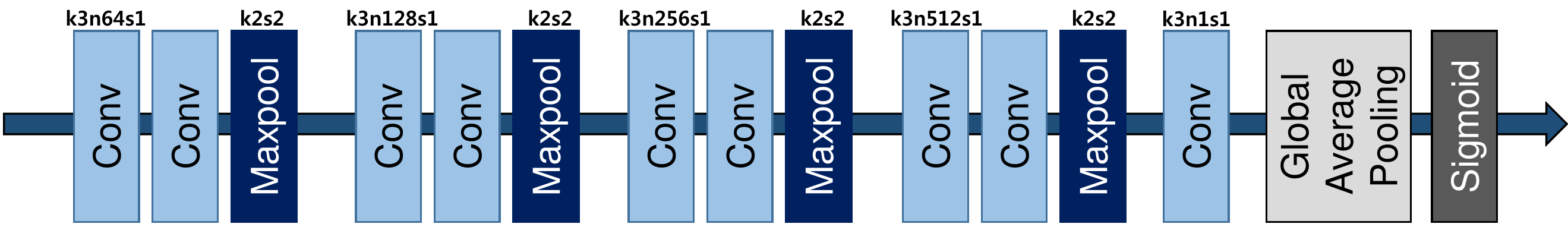}
	\end{center}
	\caption{Our NMD network architecture.}
	\label{fig:nmd}
\end{figure}

For the training, we start from $\alpha=0.5$ and $\sigma=0.1$. We update both hyper-parameters according to the average of 10 validation accuracies (AVA). When it reaches above $95 \%$, we update $\alpha$ and $\sigma$ following the rules below:
\begin{align}
\textbf{if} &\text{~AVA of~}\alpha \geq 0.95 \textbf{~then}\\
&~~~~\alpha \leftarrow \alpha + 0.1\\
\textbf{if} &\text{~AVA of~}\sigma \geq 0.95 \textbf{~then}\\
&~~~~\sigma \leftarrow 0.8 \times \sigma .
\end{align}
We stop training with the final $\alpha$ and $\sigma$ equal to $0.8$ and $0.0044$, respectively.

\section{Natural and Realistic Super-Resolution}
\label{sec:MODEL}
In this section, we explain the proposed natural and realistic super-resolution (NatSR) generator model and the training loss function.

\subsection{Network Architecture}

\begin{figure*}[t]
	\begin{center}
		\includegraphics[width=0.9\linewidth]{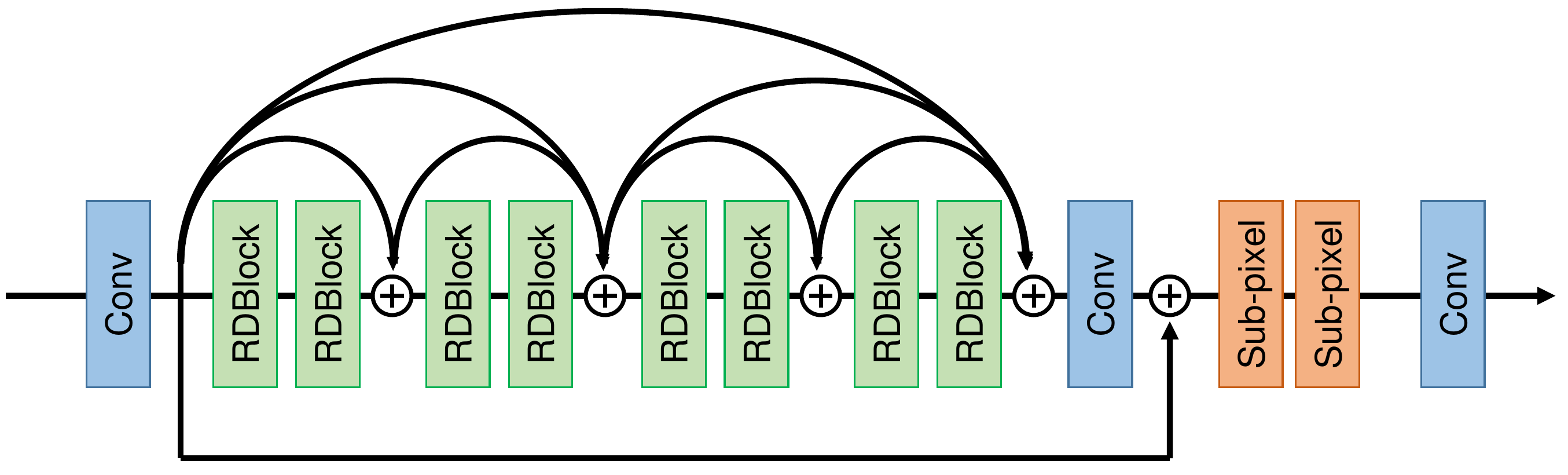}
	\end{center}
	\caption{Our NatSR network architecture. We adopt fractal residual learning
	for mid- and long-path skip connection and employ the residual dense block (RDBlock)
	for short-path connection.}
	\label{fig:fractal}
\end{figure*}

\begin{figure}[t]
	\begin{center}
		\includegraphics[width=0.8\linewidth]{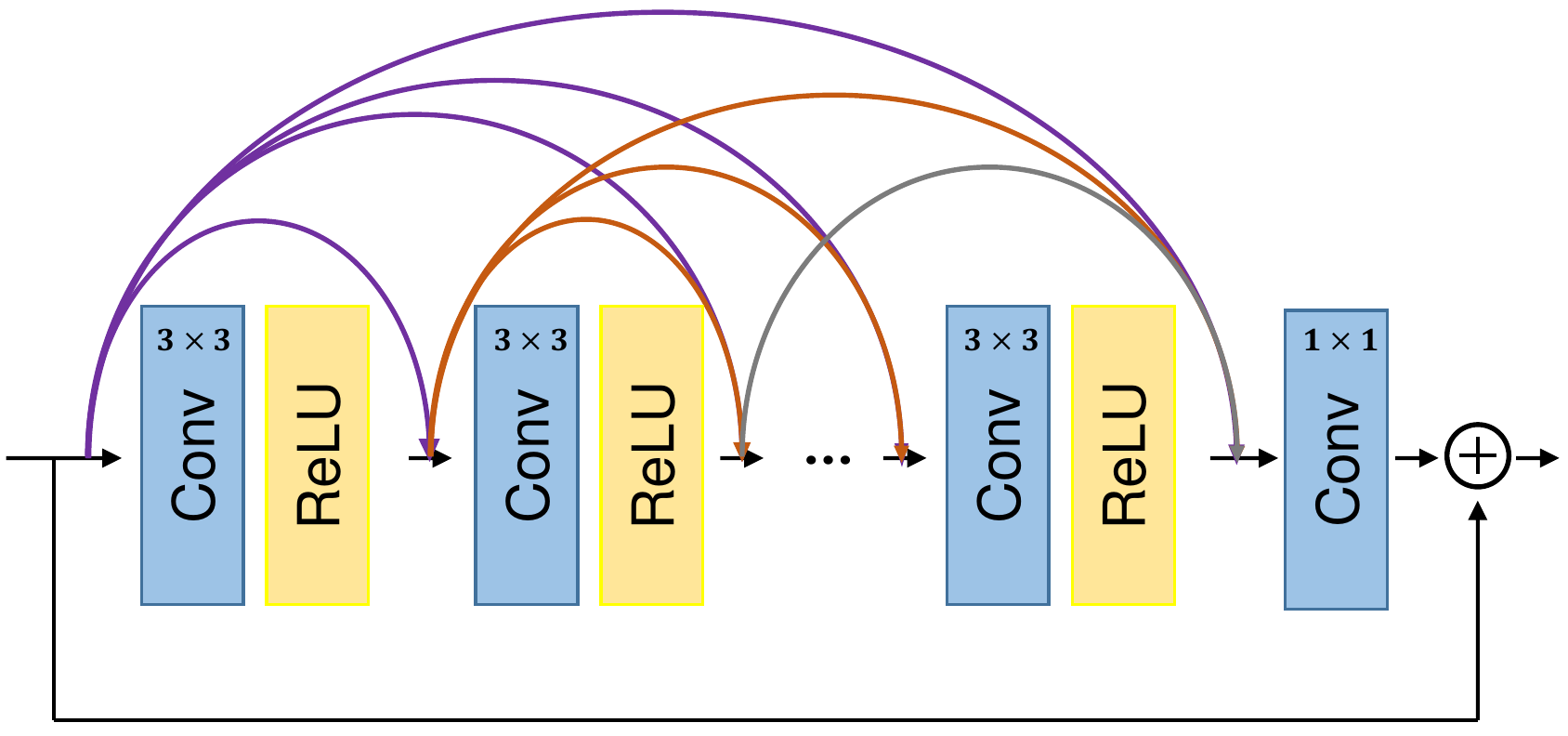}
	\end{center}
	\caption{Residual Dense Block (RDBlock) that we employ for our NatSR.}
	\label{fig:RDBlock}
\end{figure}

The overall architecture of our NatSR is shown in \figurename{~\ref{fig:fractal}}, which takes the $I_{LR}$ a the input and generates the SR output. As shown in the figure, our network is based on residual learning, which has long been used as a basic skill to mitigate the degradation problem in very deep networks. Typically, two types of residual learnings are used: local residual learning (LRL) which bypasses the input to the output in a local range \cite{ResNet}, and global residual learning (GRL) which provides the skip-connection between the input and the output in a global scale of the network \cite{VDSR}.
Former approaches \cite{VDSR, DWSR} have shown that learning the sparse features is much more effective than learning the pixel domain values directly. Hence, recent models adopt both local residual learning (short-path) and global residual learning (long-path) \cite{SRGAN, EDSR, RDN}. 

Inspired by former studies, we adopt a connection scheme shown in 
\figurename{~\ref{fig:fractal}}, named as fractal residual learning (FRL) structure in that the connection has a fractal pattern. Also, as a basic building block of our NatSR, we employ the residual dense block (RDBlock) \cite{RDN} shown in \figurename{~\ref{fig:RDBlock}}, and adopt the residual scaling \cite{EDSR} in our RDBlock. By using the FRL and RDBlock, all from short- to long-path skip-connection can be employed. 

As a discriminator for GAN, we apply a similar network architecture as NMD. Instead of using only convolution layers, we adopt spectral normalization \cite{Spectral} to make the discriminator satisfy Lipschitz condition. Also, we use strided convolutions instead of max-pooling layers. Specific architecture details are provided in the \textit{supplementary material}.

\subsection{Training Loss Function}
\subsubsection{Reconstruction Loss}
To model the $p(I_{HR}|I_{LR})$, we adopt the pixel-wise reconstruction loss, specifically the mean absolute error (MAE) between the ground-truths and the super-resolved images:
\begin{equation}
\mathcal{L}_{\text{Recon}}=\mathbb{E}[||I_{HR}-I_{SR}||_{1}],
\end{equation}
where $I_{SR}$ denotes the super-resolved output.
Although all the perception-oriented models apply perceptual losses, we do not adopt such losses, because it is found that the perceptual loss causes undesirable artifacts in our experiments. To boost high-frequency details, we instead use our NMD as a solution.

\subsubsection{Naturalness Loss}
We design the naturalness loss based on our pre-trained natural manifold discriminator (NMD). To concentrate the target manifold within the natural manifold, the output of NMD should be nearly $1$. We may use the loss as a negative of the sigmoid output, but we use its log-scale to boost the gradients:
\begin{equation}
\mathcal{L}_{\text{Natural}}=\mathbb{E} [-\log(D_{NM}(I_{SR}))]
\end{equation}
where $D_{NM}(\cdot)$ denotes the output sigmoid value of NMD.

\subsubsection{Adversarial Loss}
As it is well-known that GANs are hard to train and unstable, there have been lots of variations of GANs \cite{EBGAN, WGAN, WGAN-GP, LSGAN, RaGAN}. Recently, GAN with relativistic discriminator has been proposed \cite{RaGAN}, which shows quite robust results with standard GAN \cite{GAN} in generating fake images in terms of Fr\'{e}chet Inception Distance \cite{FID}. Thus, we employ RaGAN for our adversarial training, which is described as:
\begin{align}
\mathcal{L}_{G}&= - \mathbb{E}_{x_r\sim\mathbb{P}_{r}}[\log(\tilde{D}(x_r))]-\mathbb{E}_{x_f\sim\mathbb{P}_{g}}[\log(1-\tilde{D}(x_f))]\\
\mathcal{L}_{D}&= - \mathbb{E}_{x_f\sim\mathbb{P}_{g}}[\log(\tilde{D}(x_f))]-\mathbb{E}_{x_r\sim\mathbb{P}_{r}}[\log(1-\tilde{D}(x_r))],
\end{align}
where $\mathbb{P}_r$ and $\mathbb{P}_g$ are distributions of HR and SR respectively, $x_r$ and $x_f$ mean real and fake data respectively, and
\begin{align}
\tilde{D}(x_r)=\text{sigmoid}(C(x_r)-\mathbb{E}_{x_f\sim\mathbb{P}_g}[C(x_f)])\\
\tilde{D}(x_f)=\text{sigmoid}(C(x_f)-\mathbb{E}_{x_r\sim\mathbb{P}_r}[C(x_r)])
\end{align}
where $C(\cdot)$ denotes the output logit of discriminator.
In our case, the motivation of RaGAN discriminator is to measure ``the probability that the given image is closer to real HR images than the generated SR images on average.''

\subsubsection{Overall Loss}

The overall loss term to train our NatSR is defined as the weighted sum of loss terms defined above:
\begin{equation}
\mathcal{L}=\lambda_1 \mathcal{L}_{\text{Recon}}+\lambda_2 \mathcal{L}_{\text{Natural}}+\lambda_3 \mathcal{L}_{G}.
\end{equation}
As our baseline, we train the distortion-oriented model where $\lambda_2 = \lambda_3 = 0 $, which means that the overall loss is just the reconstruction loss $\mathcal{L}_\text{Recon}$.  We name our baseline model as fractal residual super-resolution network (FRSR).
For our NatSR which is perception-oriented, we use the full loss above with
$\lambda_1=1$, $\lambda_2=10^{-3}$ and $\lambda_3=10^{-3}$.

\section{Discussion and Analysis}
\label{fig:analysis}

\subsection{Effectiveness of Proposed Discriminator}
To demonstrate the meaning and effectiveness of our NMD,
we test the NMD scores for the perception-oriented methods such as SRGAN variants \cite{SRGAN}, EnhanceNet, NatSR, and also for the distortion-oriented methods including our FRSR. 
\tablename{~\ref{table:NMD}} shows the results on BSD100 \cite{B100}, where the NMD is designed to output score 1 when the input image is close to the natural original image, and output lower score when the input is blurry or noisy. We can see that previous perception-oriented methods score between $0$ and $1$ which means that they lie near the boundary of the natural and unnatural manifold in our LR-HR model. Also, the original HR scores $1$ and bicubic interpolation scores $0$, which means that our NMD discriminates HR and LR with high confidence. Additionally, SRResNet, EDSR, and our FRSR, which are distortion-oriented, score almost $0$. We may interpret the result that the distortion-oriented methods produce the image which also lie on the blurry manifold. On the other hand, our NatSR results in the scores close to $1$ which is much higher than the other perception-oriented algorithms.
In summary, it is believed that
our model of natural manifold and NMD are reasonable, and the NMD well discriminates the natural and unnatural manifold.

\begin{table}
	\begin{center}
		\begin{tabular}{|l|c|}
			\hline
			Method & NMD Score \\
			\hline\hline
			HR & $1.000 \pm 0.001$ \\
			Bicubic & $0.000 \pm 0.000$ \\
			\hline
			SRResNet & $0.032 \pm 0.009 $\\
			EDSR & $0.043 \pm 0.012$ \\
			FRSR (Ours) & $0.044 \pm 0.011$\\
			\hline
			SRGAN-MSE & $0.755 \pm 0.063$\\
			SRGAN-VGG22 &  $0.584 \pm 0.202$\\
			SRGAN-VGG54 & $0.832 \pm 0.109$ \\
			EnhanceNet-PAT & $0.367 \pm 0.095$\\			
			NatSR (Ours) & $ 1.000 \pm 0.000$\\
			\hline
		\end{tabular}
	\end{center}
	\caption{Results of NMD score.}
	\label{table:NMD}
\end{table}

\subsection{Study on the Plausibility of SR Images}

As we approach the SISR by interpreting the input and output images in our LR-HR space model, we analyze the plausibility of super-resolved images of various methods according to our model. The super-resolved images must lie on the set $V$ in \figurename{~\ref{fig:manifold}}, which means that the downsampling of a super-resolved image must be in the LR space, \textit{i.e.}, it must be similar to the input LR image as
\begin{equation}
I_{LR} \approx h(I_{SR})^\downarrow.
\end{equation}

For the analysis, we show the RGB-PSNR between $h(I_{SR})^\downarrow$ and $I_{LR}$ in \tablename{~\ref{table:Feasibility}} which are tested on Set5 \cite{Set5}. The results are in the ascending order of SRGAN, EnhanceNet, and our NatSR. Even though we do not give any constraints on the LR space, our NatSR results mostly lie on the feasible set $V$. On the other hand, SRGAN result is about $36$ dB, which means that the SRGAN barely reflects the LR-HR properties. 

\begin{table}
	\begin{center}
		\begin{tabular}{|l|c|}
			\hline
			Method & RGB-PSNR (dB) \\
			\hline\hline
			SRGAN & $36.16$\\
			ENet-PAT & $41.65$\\			
			NatSR& $45.94$\\
			\hline
		\end{tabular}
	\end{center}
	\caption{Results of RGB-PSNR between LR input and downsampled SR image in LR domain.}
	\label{table:Feasibility}
\end{table}

%

\section{Experimental Results}
\label{sec:Results}
\subsection{Implementation details}
We train both NMD and NatSR (including FRSR) with recently released DIV2K \cite{NTIRE} dataset which consists of high-quality (2K resolution) $800$ training images, $100$ validation images, and $100$ test images. The size of the input LR patch is set to $48\times48$, and we only train with scaling factor $\times 4$. ADAM optimizer \cite{ADAM} is used for training with the initial learning rate of $2\times10^{-4}$, and halved once during the training. We implement our code with Tensorflow \cite{TF}.
For the test, we evaluate our model with famous SISR benchmarks: Set5 \cite{Set5}, Set14 \cite{Set14}, BSD100 \cite{B100}, and Urban100 \cite{Self-Exemplar}. 

\subsection{Evaluation Metrics and Comparisons}
For the evaluation of distortion-oriented models, popular FR-IQA (full reference image quality assessment), PSNR and SSIM (structure similarity) \cite{SSIM} are used. But since these measures are not appropriate for measuring the quality of perceptual models, we use one of the recently proposed NR-IQA (no reference image quality assessment)  called NQSR \cite{NQSR} which is for SISR and well-known for Ma \etal's score. Additionally, another NR-IQA, NIQE \cite{NIQE} is used to measure the naturalness of images. The higher NQSR and the lower NIQE mean the better perceptual quality. However, it is questionable whether so many variants of NR-IQA methods perfectly reflect the human perceptual quality. Hence, we need to use the NR-IQA results just for rough reference. 

We compare our FRSR with other distortion-oriented methods such as LapSRN, SRDenseNet, DSRN, and EDSR \cite{LapSRN, SRDenseNet, DSRN, EDSR}, and
compare our NatSR with other perception-oriented ones such as SRGAN, ENet, and SFT-GAN \cite{SRGAN, EnhanceNet, SFT-GAN} (We denote SRGAN-VGG54 as SRGAN and EnhanceNet-PAT as ENet for short).

\subsection{FR-IQA Results}

\begin{table*}
	\begin{center}
			\resizebox{\linewidth}{!}{
		\begin{tabular}{|c|c|c|c|c|c|c|c||c|c|c|c||}
			\hline
			\rule[-1ex]{0pt}{3.5ex}
			Dataset & Scale & Bicubic & LapSRN & SRDenseNet & DSRN & EDSR & FRSR & SRGAN & ENet & NatSR \\
			\hline\hline
			\rule[-1ex]{0pt}{3.5ex}
			Set5    & 4 & 28.42/0.8104 & 31.54/0.8850 & 32.02/0.8934 & 31.40/0.8830 & \textcolor{red}{32.46/0.8976}  & \textcolor{blue}{32.20/0.8939} & 29.41/0.8345 &28.56/0.8093 & 30.98/0.8606\\
			\hline
			\rule[-1ex]{0pt}{3.5ex}
			Set14   & 4 & 26.00/0.7027 & 28.19/0.7720 & 28.50/0.7782 & 28.07/0.7700 & \textcolor{red}{28.71/0.7857} & \textcolor{blue}{28.54/0.7808} & 26.02/0.6934 &25.67/0.6757
			 &27.42/0.7329\\
			\hline
			\rule[-1ex]{0pt}{3.5ex}
			BSD100  & 4 & 25.96/0.6675 & 27.32/0.7280 & 27.53/0.7337 & 27.25/0.7240 & \textcolor{red}{27.72/0.7414} & \textcolor{blue}{27.60/0.7366} & 25.18/0.6401 &24.93/0.6259 & 26.44/0.6827\\
			\hline
			\rule[-1ex]{0pt}{3.5ex}
			Urban100& 4 & 23.14/0.6577 & 25.21/0.7560 & 26.05/0.7819 & 25.08/0.7470 & \textcolor{red}{26.64/0.8029} & \textcolor{blue}{26.21/0.7904} & - &23.54/0.6926 &25.46/0.7602\\
			\hline\hline
			Parameters & 4 & - & 0.8 M & 2.0 M & 1.2 M & 43 M & 4.9 M & 1.5 M & 0.8 M & 4.9 M \\
			\hline
		\end{tabular}
	}
	\end{center}
	\caption{FR-IQA results. The average PSNR/SSIM values on benchmarks. \textcolor{red}{Red} color indicates the best results, and the \textcolor{blue}{blue} indicates the second best.}
	\label{table:PSNR}
\end{table*}

In this subsection, we discuss the distortion-oriented methods and their results. The overall average PSNR/SSIM results are listed in \tablename{~\ref{table:PSNR}}, which shows that our FRSR shows comparable or better results compared to the others. The EDSR \cite{EDSR} shows the best result, however, considering the number of parameters shown in the last row of \tablename{~\ref{table:PSNR}}, our FRSR is also a competent method.
As a sub-experiment, we also evaluate the FR-IQA results on the perception-oriented methods. Of course, the results are worse than the distortion-oriented algorithms, sometimes even worse than the bicubic interpolated images. Nonetheless, ours are slightly nearer to the original image in the pixel-domain than the SRGAN and EnhanceNet.

\subsection{NR-IQA Results}
\begin{figure}[t]
	\begin{center}
		\begin{subfigure}[t]{0.48\linewidth}
			\centering
			\includegraphics[width=1\columnwidth]{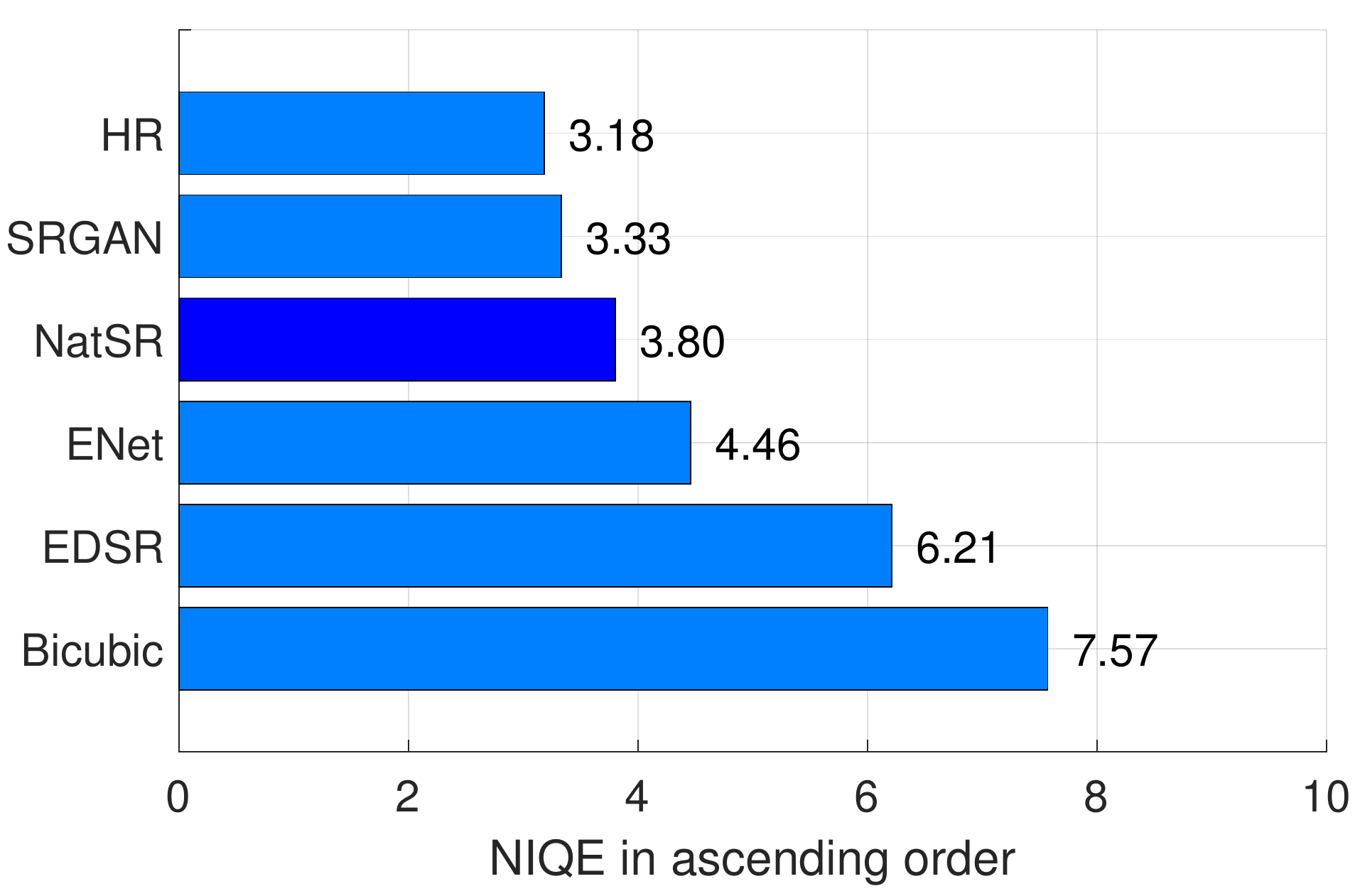}
			\caption{NIQE in ascending order.}
			\label{fig:NIQE}
		\end{subfigure}
		\begin{subfigure}[t]{0.48\linewidth}
			\centering
			\includegraphics[width=1\columnwidth]{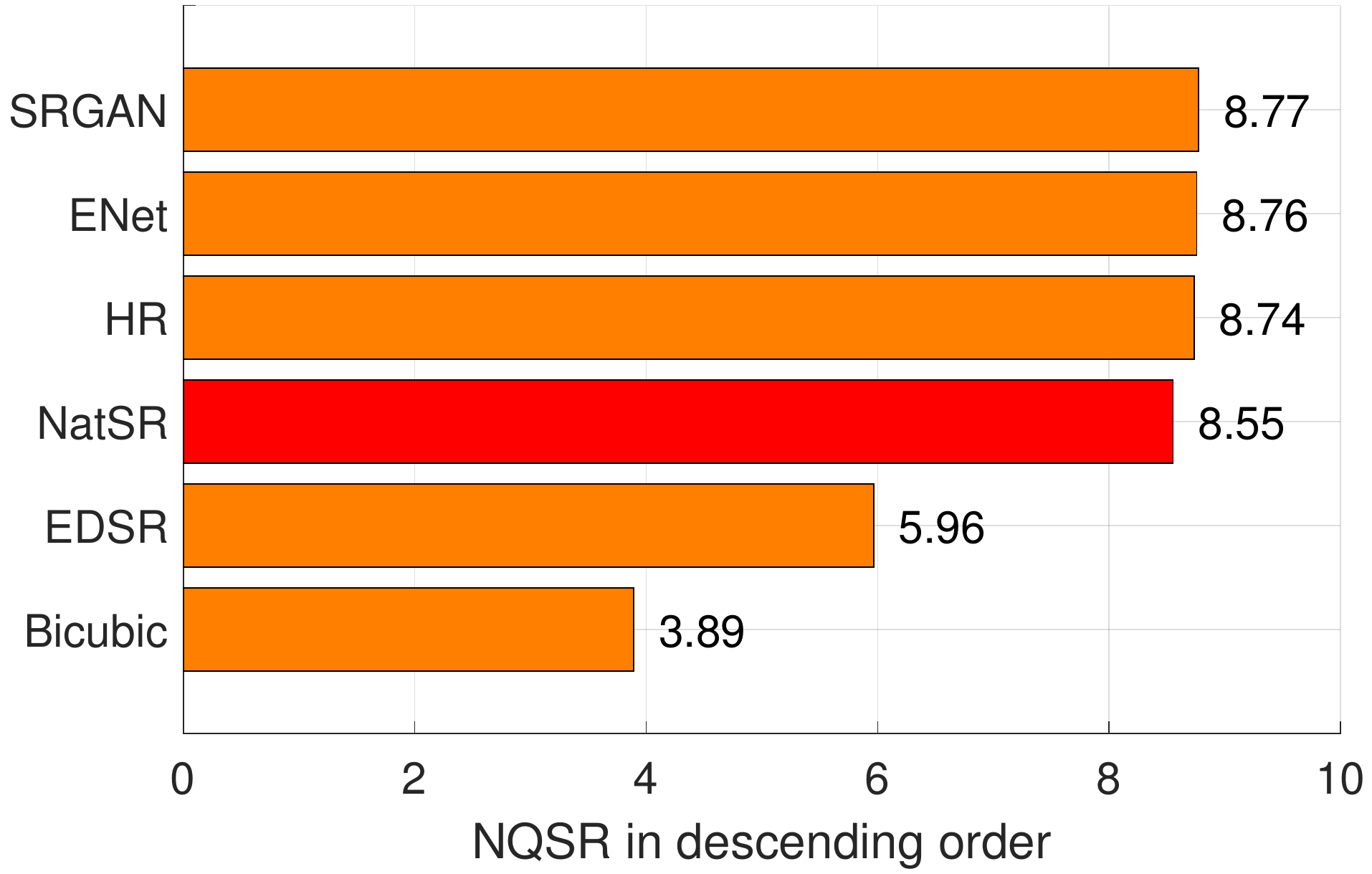}
			\caption{NQSR in descending order.}
			\label{fig:NQSR}
		\end{subfigure}

	\end{center}
	\caption{NR-IQA results in the sorted order (left: NIQE \cite{NIQE}, and right: NQSR \cite{NQSR}). The best is at the top and the worst is at the bottom.Our NatSR result is highlighted with darker color.}
	\label{fig:NR-IQA}
\end{figure}

We assess the methods with the NR-IQAs and the results are summarized in \figurename{~\ref{fig:NR-IQA}}, which shows the average NIQE and NQSR tested with BSD100. As can be observed, our NatSR is not the best but yields comparable measures to other perception-oriented methods and the original HR. As expected, one of the state-of-the-art distortion-oriented methods, EDSR scores the worst in both metrics except for the bicubic interpolation. For NIQE, besides the ground-truth HR, SRGAN scores the best. Our NatSR scores the second best for this metric. For NQSR, SRGAN scores the best among all methods including the HR. Our NatSR ranks lower than SRGAN and ENet, but the scores of all the methods including the HR show a slight difference. Although the NatSR is not the best in both scores, we believe NatSR shows quite consistent results to human visual perception as shown in Figures~\ref{fig:001} and \ref{fig:002}, by suppressing the noisy and blurry outputs through the NMD cost.

\begin{figure*}[t]
	\begin{center}
	\begin{subfigure}[t]{0.32\linewidth}
		\centering
		\includegraphics[width=1\columnwidth]{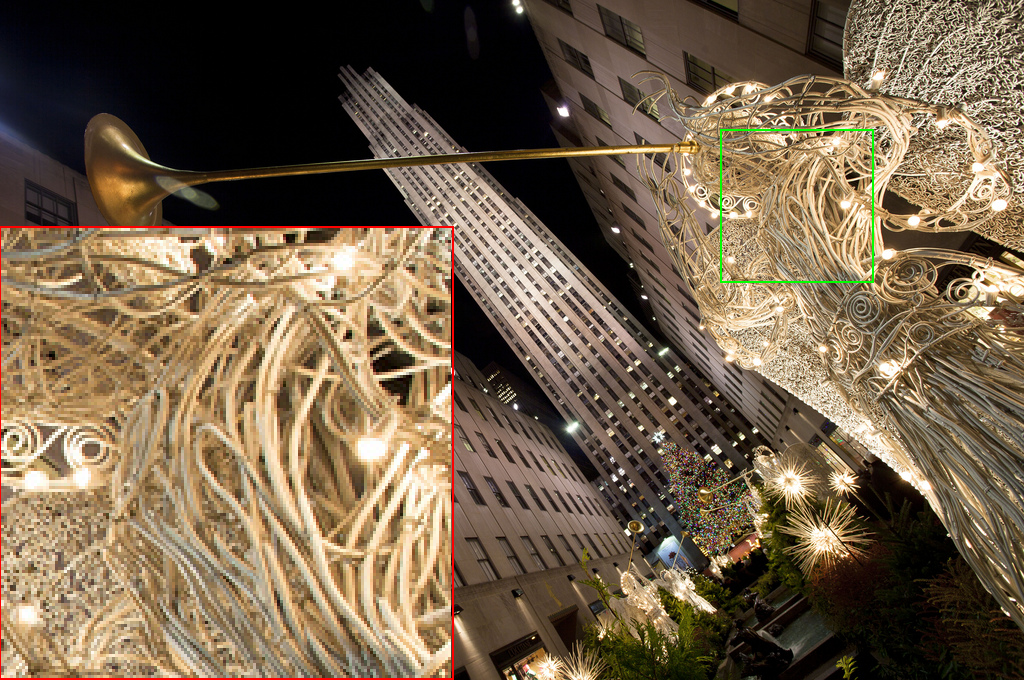}
		\caption*{HR}
	\end{subfigure}
	\begin{subfigure}[t]{0.32\linewidth}
		\centering
		\includegraphics[width=1\columnwidth]{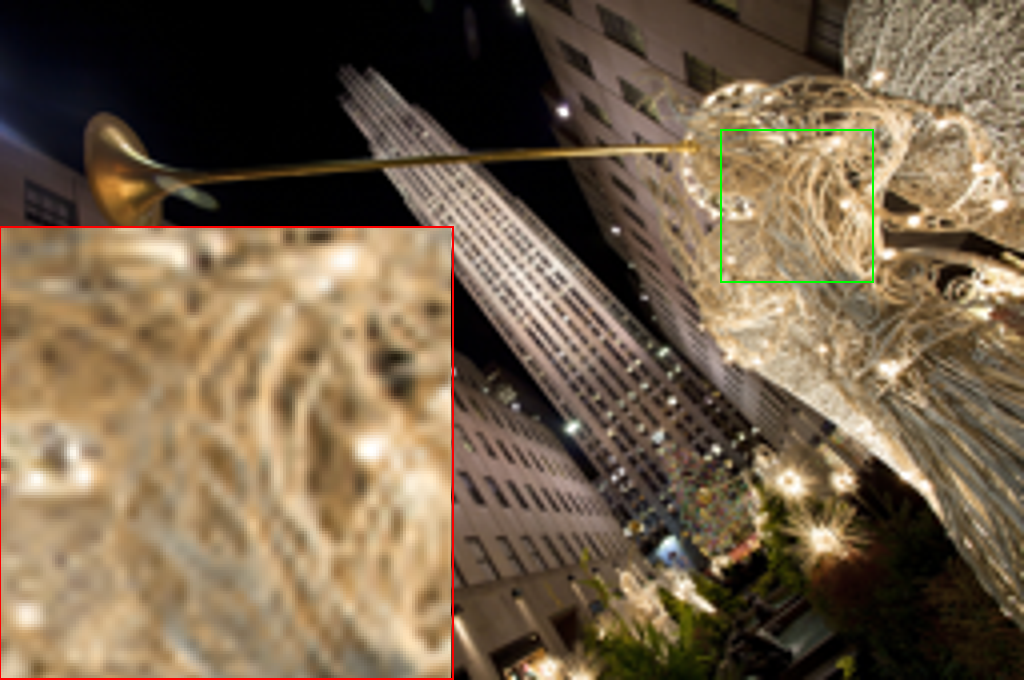}
		\caption*{Bicubic}
	\end{subfigure}
		\begin{subfigure}[t]{0.32\linewidth}
		\centering
		\includegraphics[width=1\columnwidth]{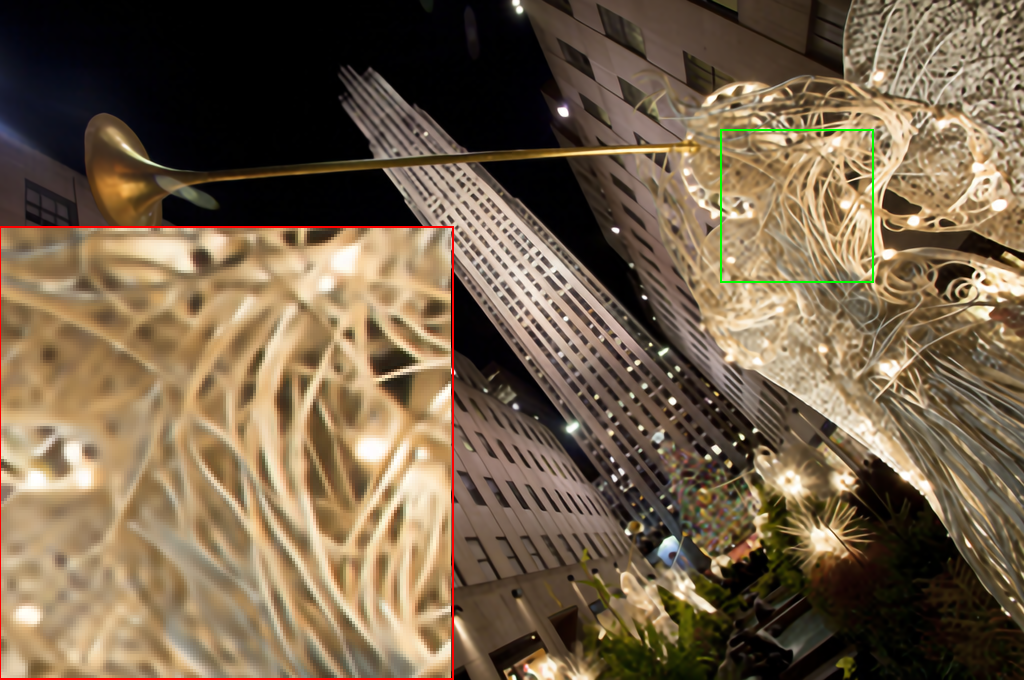}
		\caption*{EDSR}
	\end{subfigure}
	\begin{subfigure}[t]{0.32\linewidth}
		\centering
		\includegraphics[width=1\columnwidth]{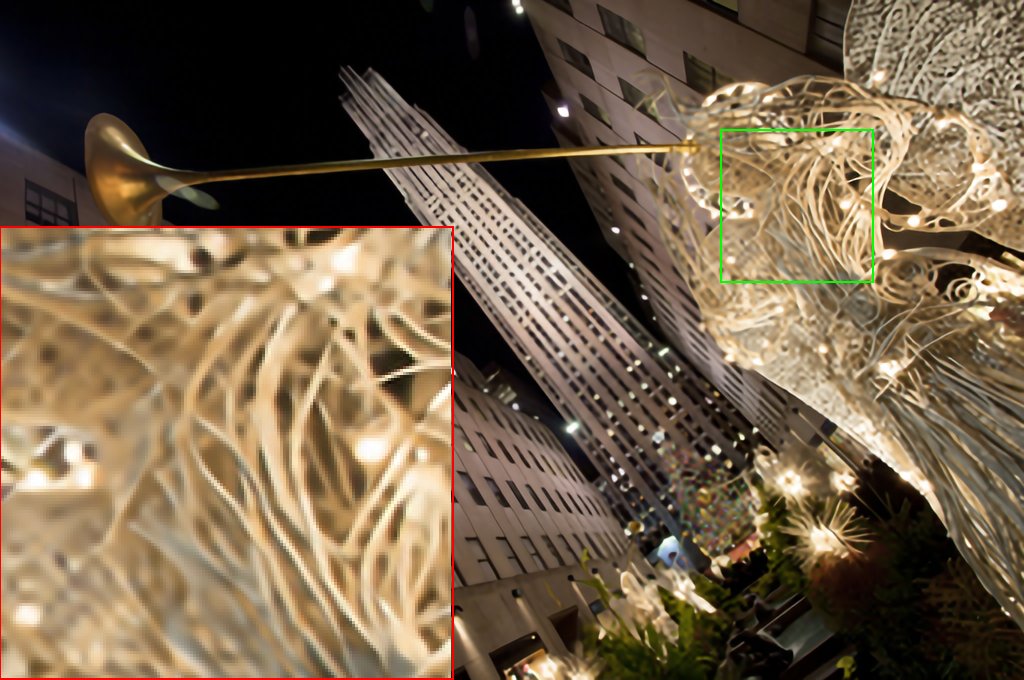}
		\caption*{FRSR (Ours)}
	\end{subfigure}
		\begin{subfigure}[t]{0.32\linewidth}
		\centering
		\includegraphics[width=1\columnwidth]{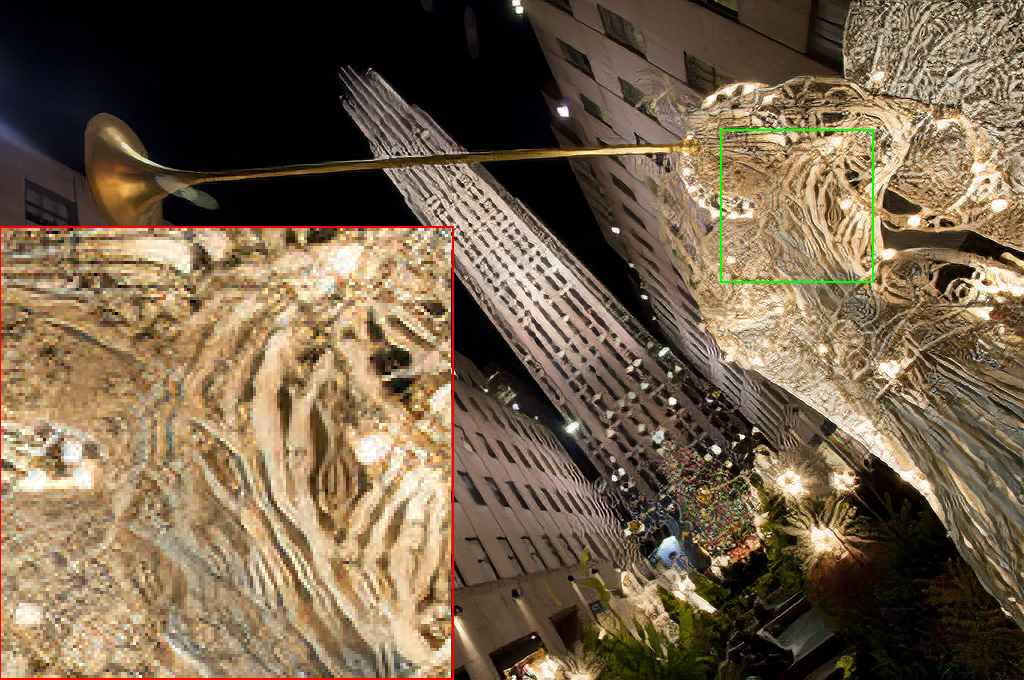}
		\caption*{ENet}
	\end{subfigure}
	\begin{subfigure}[t]{0.32\linewidth}
		\centering
		\includegraphics[width=1\columnwidth]{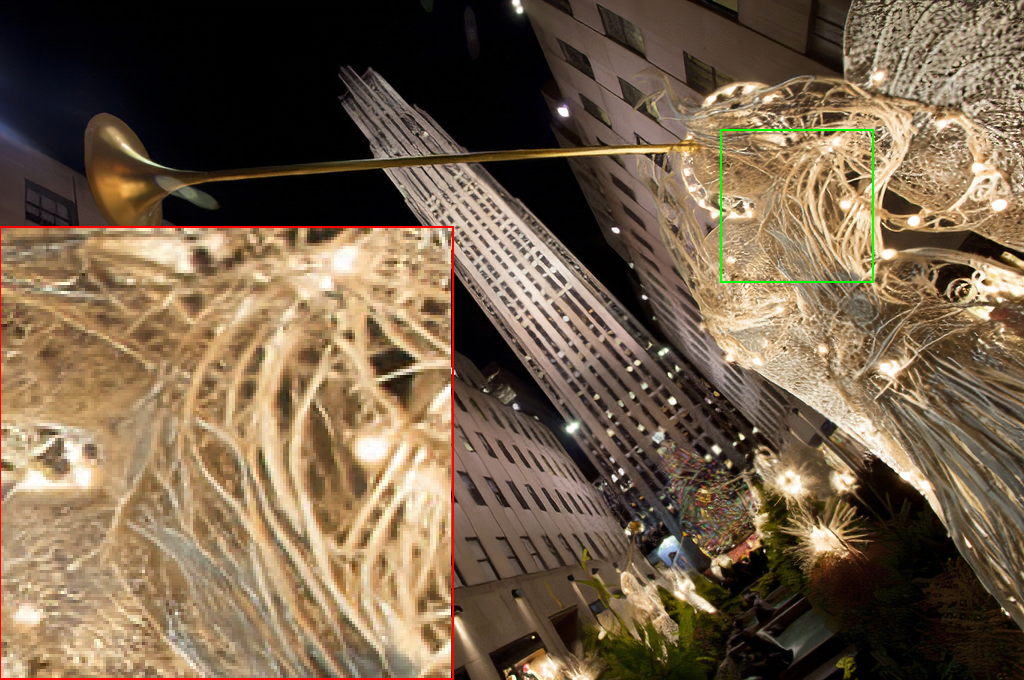}
		\caption*{NatSR (Ours)}
	\end{subfigure}

	\end{center}

	\caption{Visualized results on ``img031'' of Urban100.}
	\label{fig:002}
\end{figure*}


\section{Subjective Assessments}

\subsection{Mean Opinion Score (MOS)}
To better assess the perceptual quality of several results, we conduct a mean opinion score (MOS) test with DIV2K validation set \cite{NTIRE}. For the fair comparison with recent perception-oriented methods, SFT-GAN \cite{SFT-GAN} is evaluated with proper semantic segmentation mask to generate the best performance. The details are in \emph{supplementary material}.

%

\subsection{Visual Comparisons}
We visualize some results in \figurename{~\ref{fig:001}}, \ref{fig:002}}. As shown in Figure~\ref{fig:001}, our NatSR shows the least distortion compared to other perception-oriented methods. Also, Figure~\ref{fig:002} shows that distortion-oriented methods show blurry results while perception-oriented ones show better image details. However, ENet produces unnatural cartoony scenes, and SFT-GAN fails to produce natural details in buildings. More results can be found in \emph{supplementary material}. 

\section{Conclusion}
In this paper, we have proposed a new approach for SISR which hallucinates natural and realistic textures. First, we start from the modeling of LR-HR space and SISR process. From this work, we developed a CNN-based natural manifold discriminator, which enables to narrow the target space into the natural manifold. We have also proposed the SR generator based on the residual dense blocks and fractal residual learning. The loss function is designed such that our network works either as a distortion-oriented or perception-oriented model. From the experiments, it is shown that our distortion-oriented network (FRSR) shows considerable gain compared to the models with similar parameters. Also, our perception-oriented network (NatSR) shows perceptually plausible results compared to others. We expect that with deeper and heavier network for generating better super-resolved images and also with better classifier as NMD, our method would bring more naturalness and realistic details. 
The codes are publicly available at \url{https://github.com/JWSoh/NatSR}.

\vspace{-1.5 em}
\paragraph{Acknowledgments}
This research was financially the Ministry of Trade,
Industry, and Energy (MOTIE), Korea, under the ``Regional Specialized Industry Development Program(R\&D, P0002072)''
supervised by the Korea Institute for Advancement of Technology (KIAT).

\end{document}